\documentstyle[preprint,aps,prd]{revtex}

\begin{document}
 
\title {{\rm \hfill Imperial/TP/97-98/67 \vspace{0.8cm}}\\
Energy-momentum tensor for scalar fields coupled 
to the dilaton in two dimensions}    

\author{Fernando C.\ Lombardo \footnote{Electronic address: 
lombardo@df.uba.ar} and Francisco D.\ Mazzitelli 
\footnote{Electronic address: fmazzi@df.uba.ar}}

\address{{\it
Departamento de F\'\i sica, Facultad de Ciencias Exactas y Naturales\\ 
Universidad de Buenos Aires - Ciudad Universitaria, 
Pabell\' on I\\ 
1428 Buenos Aires, Argentina}}

\author{Jorge G.\ Russo \footnote{Electronic address: jrusso@ic.ac.uk}}

\address{{\it Theoretical Physics Group, Blackett Laboratory,\\
Imperial College, London SW7 2BZ, U.K.}}

\maketitle

\begin{abstract}
We clarify some issues related to the evaluation of the mean value of
the energy-momentum tensor for quantum scalar fields coupled to the
dilaton field in two-dimensional gravity. Because of this coupling,
 the  energy-momentum tensor for the matter 
is not conserved and therefore it is not
determined by the trace anomaly. We discuss different approximations 
for the calculation of the energy-momentum tensor and show how to obtain 
the correct amount of Hawking radiation. We also compute cosmological 
particle creation and quantum corrections to the Newtonian potential.

\end{abstract}
\newpage

\section{Introduction}

In semiclassical and quantum gravity it is of interest to compute the 
backreaction of quantum fields on the spacetime geometry. 
Given that a complete four-dimensional calculation is obviously a 
complicated problem,
one may first try to investigate it in  two-dimensional models, where 
non-spherical degrees of freedom are truncated.
The two-dimensional model of Callan et al \cite{CGHS} consists of a 
 metric coupled to  a dilaton field $\phi $ and conformal
matter fields $f$. The action is given by
\begin{equation} 
S = \int d^2x \sqrt{g}\left\{{e^{-2\phi}\over{16\pi}}\left[R +
4 (\partial\phi)^2 + 4\Lambda^2 \right]- {1\over{2}} (\partial f)^2\right\}.
\label{cghs}
\end{equation}
By virtue of the  conformal symmetry of the classical action, 
quantum 
effects of the matter fields are essentially given by the trace anomaly
$\langle T_a^a\rangle =R/24\pi$.
The mean value of the energy-momentum tensor is determined by this anomaly
and the conservation law $\langle T^{ab}\rangle_{;b} =0$. By including 
$\langle T_{ab}\rangle $ in the equations of motion, it is possible to 
study backreaction effects on the spacetime geometry.

In order to make contact with four dimensions, one may consider 
the usual Einstein-Hilbert action and minimally 
coupled scalar fields 
\begin{equation}
S = \int d^4x \sqrt{g^{(4)}}\left[{1\over{16\pi}} R^{(4)}
- {1\over{2}} (\partial^{(4)}f)^2\right]\ .
\label{eh4dact}
\end{equation}
For spherically symmetric configurations
\begin{equation}
ds^2=g_{\mu\nu}dx^{\mu}dx^{\nu}=g_{ab}(x^a)dx^{a}dx^{b}+ e^{-2\phi (x^a)}
(d\theta^2+
\sin^2\theta \ d\varphi^2)\ ,~f=f(x^a)\ ,~a,b=0,1\ ,
\label{aaa}
\end{equation}
the action reduces to
\begin{equation}
S = \int d^2x \ \sqrt{g} e^{-2\phi}\left[ {1\over{16\pi}}\left( R +
2 (\partial\phi)^2 + 2 e^{2\phi}\right) - {1\over{2}} (\partial f)^2\right].
\label{2dact}
\end{equation}
Unlike model (\ref{cghs}), matter fields originating from four dimensions
are coupled to the dilaton field.

Similarly, starting with non-minimally coupled scalar fields with action
\def\ha{ {\textstyle{{1\ov 2}}} }
\def\ov{\over}
\def\del{\partial }

\begin{equation}
S_{\rm matter} =-\ha \int d^4x \sqrt{g^{(4)}}\left[ (\partial^{(4)}f)^2+\xi  
R^{(4)} f^2
 \right]\ ,
\label{bbb}
\end{equation}
one gets the following action upon reduction:
\begin{equation}
S_{\rm matter} =-\ha \int d^2 x \sqrt{g} e^{-2\phi}\left[ (\partial f)^2+\xi  
f^2\big(
R^{(2)} +4\Box\phi - 6 (\del \phi )^2+2 e^{2\phi }\big)
 \right]\ ,
\label{cbb}
\end{equation}
which, in terms of $\psi=e^{-\phi} f$, reads
\begin{equation}
S_{\rm matter} =-\ha  \int d^2 x \sqrt{g} \left[(\partial \psi)^2+V \psi^2
\right]\ ,
\label{ccc}
\end{equation}
with
\begin{equation}
V=\xi R^{(2)} +(4\xi -1) \Box\phi +(1- 6\xi) (\del \phi )^2+2 \xi e^{2\phi }\ .
\label{ddd}
\end{equation}
Special cases are $\xi=0$ and $\xi=1/6$.
For $\xi =1/6$, the action is conformal invariant in four dimensions, i.e. 
invariant
under $g_{\mu\nu}\to e^{2\sigma(x)} g_{\mu\nu}$ and $f \to e^{-\sigma (x)} f$.
From two-dimensional viewpoint, this implies (cf. 
Eq.~(\ref{ddd})~) $g_{ab}\to e^{2\sigma(x)} g_{ab}
\ ,\ \ 
\phi\to \phi-\sigma$, and $\psi\to \psi $ (or $f \to e^{-\sigma (x)} f$~). 
The  matter action 
in (\ref{2dact}), corresponding to $\xi =0$,
is conformal invariant in two dimensions, i.e. under the transformation
$g_{ab}\to e^{2\sigma(x)} g_{ab}\ ,\ \ 
\phi\to \phi $, and $f \to f$. For any other $\xi\neq 0,1/6 $, there is no 
invariance involving Weyl scalings in the two-dimensional model.

\phantom{\cite{muk,hb1,kummer,odin,dowk} }

Let us now consider the model (\ref{2dact}). Due to  the conformal symmetry,
 the trace 
of the energy-momentum tensor of the scalar fields vanishes classically. There 
is, however, an anomaly at the quantum level. This anomaly has been 
computed by a number of authors  (see refs.~\cite{muk}-\cite{dowk}).
Some new, interesting effects have been discussed, including
the (anti) evaporation of Schwarzschild-de Sitter
black holes  \cite{hb2}. 
However, it has also been
claimed (based on an energy-momentum tensor obtained by using the 
conservation law)
that quantum effects due to the anomaly produce an {\it ingoing}
Hawking radiation for Schwarzschild black holes \cite{muk,kummer}. 
This seems in contradiction with the expectation that the outgoing
energy-density flux of 
Hawking radiation in four dimensions is positive definite, even in the 
$s$-wave sector. The confusion was partly clarified in a recent paper by 
Balbinot and Fabbri \cite{balbi},
who pointed out that, due to the coupling between the scalar field and the 
dilaton, 
the two-dimensional energy-momentum tensor of matter  is not conserved and 
therefore the
knowledge of the anomaly is not enough to determine the full
energy-momentum tensor. In the same paper they have also raised new puzzles 
concerning divergences in the  mean value of the energy-momentum tensor.

The aim of this paper is to clarify these puzzles and some
confusion existing in the literature
about the calculation of the energy-momentum tensor of the matter fields 
in the spherically 
reduced models.  We will compute
the effective action and the energy momentum tensor using different 
approximations, and discuss the validity of each approximation. 
It will be shown that the
energy-momentum tensor can be written as the sum of two terms:
an anomalous conserved
part and a traceless, non-conserved contribution. As we will see,
the last term is
relevant for quantum effects on black holes and cosmological spacetimes.

\section{The effective action}

At the classical level the energy-momentum tensor of the matter fields
is given by 
\begin{equation}
T_{ab}= e^{-2\phi}\left[ \partial_a f \partial_b
f - {1\over{2}} g_{ab} (\partial f)^2\right]\ .
\label{tmunu2d} 
\end{equation}
It is important to note that this energy-momentum
tensor is traceless
and  not 
conserved. Indeed, after using the classical equation of motion 
for $f$, the divergence is given by
\begin{equation}
\nabla^a T_{ab}= -{1\over{2}}\partial_a (e^{-2\phi})
(\partial f)^2.
\label{divt2d}
\end{equation}
Of course, the quantity that is conserved by Noether theorem in this 
theory is the  complete energy-momentum tensor. 

The reason why $T_{ab}$ is not conserved is also clear from the 
four-dimensional
origin of $T_{ab}$. Indeed, from 
$$
\nabla^\mu T_{\mu\nu}^{(4)}=0\ ,
$$ 
and using Eq.~(\ref{aaa}), one obtains
\begin{equation}
\nabla^a T_{ab}^{(4)}=2\del^a\phi T^{(4)}_{ab}-e^{2\phi }
(\del_b\phi T^{(4)}_{\theta\theta}
+\sin^{-2}\theta \del_b\phi T_{\varphi\varphi}^{(4)}) ,
\label{zaa}
\end{equation}
which, after using $T_{\mu\nu}^{(4)}=  \partial_\mu f \partial_\nu
f - {1\over{2}} g_{\mu\nu} (\partial f)^2 $ and $f=f(x^a)$, 
reproduces Eq.~(\ref{divt2d}).

At the quantum level, the mean value $\langle T_{ab}\rangle $ is a 
divergent quantity
that must be renormalized. In view of the above discussion, we expect that a 
covariant renormalization will 
produce a non-conserved energy-momentum tensor with a trace anomaly. To check
this, we must calculate the effective action. The matter  
action in Eq. (\ref{ccc}) can be 
written as 
\begin{equation}
S_{\psi} = -{1\over{2}}\int d^2x ~ \sqrt{g}\left[
(\partial \psi)^2 + P~\psi^2\right],
\end{equation}
where $P=(\partial\phi)^2-\Box\phi$.

The Euclidean effective action can be computed using the fact that, at the 
quantum level, the trace of the energy-momentum tensor is given by  
$T=2 g^{ab}{\delta S\over\delta g^{ab}}={1\over 24\pi}(R -6P)$ 
\cite{muk,kummer,dowk}. 
Integrating this equation we obtain
\begin{eqnarray}
S_{\rm eff} &=& -{1\over{8\pi}}
\int d^2x ~ \sqrt{g}
\int d^2y ~ \sqrt{g} \left\{{1\over{12}} R(x) {1\over{\Box}} R(y) -
 P(x) {1\over{\Box}} R(y) \right\} + S_{\rm eff}^{\rm I}\nonumber\\ 
&\equiv & S_{\rm eff}^{\rm A} + S_{\rm eff}^{\rm I}\ .
\label{niWefacc} 
\end{eqnarray}
The first term in the above equation $S_{\rm eff}^{\rm A}$ produces the 
expected
anomaly, whereas the second term is Weyl invariant and non-trivial due to the 
coupling between dilaton and  scalar fields.

Working in the conformal gauge, the invariant term can be 
written as 
\begin{equation}
e^{- S_{\rm eff}^{\rm I}}= \mbox{det} 
[-\Box_f+P_f]^{-{1\over{2}}} = {\cal N} \int {\cal D}\psi\ e^{-{1\over{2}}
\int d^2x \ \psi (-\Box_f )\psi} ~ e^{-{1\over{2}}\int d^2x\ P_f \psi^2},
\label{sefinv}
\end{equation}
where the subindex $f$ indicates that the quantity must be evaluated in a
flat metric, and ${\cal N}$ is a normalization constant.
In some previous works, 
the invariant  term was simply omitted \cite{miko}. 
A possibility is to compute it using an expansion in powers 
of $P_f$ \cite{avra}:
\begin{equation}
S_{\rm eff}^{\rm I} = \int d^2x P_f(x) D_1(x) + \int d^2x \int d^2y 
P_f(x) D_2(x,y) P_f(y)+... 
\label{ex}
\end{equation}
Comparing terms of the same order in Eqs.~(\ref{sefinv}) and (\ref{ex})
we obtain $D_1(x) = \ha G(x,x)$, $D_2(x,y) = {1\ov 4} G^2(x,y)$, where $G$ 
is the 
flat Euclidean propagator. Therefore, to second order in the expansion in 
powers of $P$,
 the effective action is given by 
\begin{equation}
S_{\rm eff}^{\rm I} = {1\over{4}}\int d^2x\int d^2y P_f(x) G^2(x,y) 
P_f(y)\ ,
\end{equation}
where we have omitted a local divergent term, which can be removed by a 
counterterm.
 
The square of the propagator $G^2$ was derived in
 \cite{kum2d}. It is given by
\begin{equation}
 G^2(p) = {1\over{2\pi}} {1\over{p^2}}\ln {p^2\over{\mu^2}}.
\label{g2}
\end{equation} 
Taking this into account, the result 
up to second order in $P_f$ is
\begin{eqnarray}
S_{\rm eff}^{\rm I} &=& -{1\over{8\pi}}\int d^2x 
\int d^2y 
 P_f(x) {1\over{\Box_f}}\ln {-\Box_f\over{\mu^2}} P_f(y)
\nonumber\\
&=& -{1\over{8\pi}}\int d^2x ~ \sqrt{g}
\int d^2y ~ 
\sqrt{g} ~ P(x) {1\over{\Box}}\ln {-\Box\over{\mu^2}} P(y)\nonumber \\
&+& {1\over{8\pi}} \int d^2x ~ \sqrt{g} \int d^2z ~ \sqrt{g}
\int d^2y ~ 
\sqrt{g} ~ P(x) {1\over{\Box}}R(z){1\over{\Box}} P(y),
\label{avra}
\end{eqnarray}
where we have performed the Fourier transform of Eq.(\ref{g2}). In the 
second line we have written the effective action in an 
explicitly covariant way using that $P_f=\sqrt g P$ and that the Green 
function ${1/\Box_f}$ is Weyl invariant. The parameter
$\mu$ is an infrared cut-off, and the effective action is $\mu$-dependent 
because we are computing perturbations around massless fields in two 
dimensions. Physical results will depend on $\mu$ in this approximation. It 
is worth noting that this calculation of $S_{\rm eff}^{\rm I}$ is valid 
up to second order in $P$, but no expansion in powers of the curvature $R$ 
has been performed; in this sense, this derivation differs from the one 
given in \cite{avra}. 

To avoid infrared divergences, in Ref. \cite{muk} $S_{\rm eff}^{\rm I}$  has 
been computed 
by assuming that the mass term in Eq.(\ref{sefinv}) is a constant. 
This approximation corresponds to neglecting the backscattering of the
geometry on the dynamics of the matter fields.
In this approximation the effective action reads
\begin{eqnarray}
S_{\rm eff}^{\rm I} &=& -{1\over{8\pi}} \int d^2x  
P_f \left( 1 - \mbox{log}{P_f\over{\mu^2}}\right) 
\nonumber \\
&=& -{1\over{8\pi}} \int d^2x ~ \sqrt{g} ~ 
P(x) \left( 1 - \mbox{log}{P\over{\mu^2}}\right)-{1\over{8\pi}} 
\int d^2x \sqrt{g}\int d^2y \sqrt{g} P(x) {1\over{\Box}} R(y)\ .
\label{mukha}
\end{eqnarray} 
The last term in Eq.~(\ref{mukha}) will cancel against a similar term in 
$S_{\rm eff}^{\rm A}$ (see Eq.~(\ref{niWefacc})).
The explicit covariant expression above has been obtained by noting 
that, in the conformal gauge, $\log(\sqrt g)= -\Box ^{-1} R $. 
As has been shown in Ref.\cite{muk},
it is possible to go beyond the no-backscattering approximation by doing 
perturbations in powers of derivatives of $P$.

In both approximations the effective action can be written as 
$S_{\rm eff}=S_{\rm eff}^{\rm A}+ S_{\rm eff}^{\rm I}$. Therefore, 
a similar decomposition holds for the energy-momentum tensor
$\langle T_{ab}\rangle =\langle T_{ab}^{\rm A}\rangle + \langle 
T_{ab}^{\rm I}\rangle $. The anomalous part is
independent of the approximation and is given by 
\begin{eqnarray}
\langle T_{ab}^{\rm A}\rangle &=& {1\over{4\pi}} 
\int d^2y \sqrt{g} \left[ \nabla_a \nabla_b - g_{ab}
\Box \right]_{(x)} P(y) {1\over{\Box}} \nonumber \\
&-& {1\over{24\pi}}  \int d^2y \sqrt{g} \left[ \nabla_a \nabla_b - 
g_{ab} \Box \right]_{(x)} R(y) {1\over{\Box}}\nonumber \\
&+& {1\over{8\pi}}\int d^2y \sqrt{g} 
\left[ g_{ab} \nabla^c \phi
\nabla_c - 2 \nabla_a \phi \nabla_b + g_{ab}(\partial \phi)^2 - 
2 \nabla_a \phi \nabla_b \phi\right]_{(x)} R(y)
{1\over{\Box}} \nonumber \\
&+& {1\over{96\pi}} \int d^2x \sqrt{g} \int d^2y \sqrt{g}~\left\{
 2 \partial_a {R(x)\over{\Box}} \partial_b {R(y)\over{\Box}}
- g_{ab} \partial^c{R(x)\over{\Box}} 
\partial_c {R(y)\over{\Box}}\right\} \nonumber \\
&-& {1\over{8\pi}} \int d^2x \sqrt{g}\int d^2y \sqrt{g}~\left\{
2 \partial_a {P(x)\over{\Box}} \partial_b {R(y)\over{\Box}}
- g_{ab} \partial^c{P(x)\over{\Box}}
\partial_c{R(y)\over{\Box}} \right\}
\label{ancon}\ .
\end{eqnarray}
Note that $\langle T_{ab}^{\rm A}\rangle$ has the correct trace anomaly and 
also contains a traceless, non-conserved part.

In the 
 approximation obtained by expanding in powers of $P$,
the non-anomalous part of the energy-momentum tensor reads, up to linear order 
in $P$, as 
\begin{eqnarray}
\langle T_{ab}^{\rm I}\rangle &=&  - {1\over{4\pi}} \int d^2y \sqrt{g} 
\left[ g_{ab}\nabla^{c} \phi
\nabla_{c} - 2 \nabla_a\phi \nabla_b +
g_{ab}(\partial \phi)^2 - 
2 \nabla_a \phi \nabla_b \phi
\right]_{(x)}\nonumber \\
&\times & \left({1\over{\Box}}\ln {-\Box\over{\mu^2}} P(y)
-\int d^2z {1\over{\Box}}R(z){1\over{\Box}} P(y)\right)
,\label{nanoool}
\end{eqnarray}
while in the no-backscattering approximation it is given by
\begin{eqnarray}
\langle T_{ab}^{\rm I}\rangle &=&  {1\over{8\pi}}\left[ g_{ab} \nabla^c \phi
\nabla_c - 2 \nabla_a \phi \nabla_b +
g_{ab}(\partial \phi)^2 - 
2 \nabla_a \phi \nabla_b \phi
\right]~\mbox{log}{P\over{\mu^2}}
\nonumber \\
&-& {1\over{8\pi}}g_{ab} 
P ~ - {1\over{4\pi}} 
\int d^2y \sqrt{g} \left[ \nabla_a \nabla_b - g_{ab}
\Box \right]_{(x)} P(y) {1\over{\Box}} \nonumber \\
&-& {1\over{8\pi}}\int d^2y \sqrt{g} \left[ g_{ab} \nabla^c \phi
\nabla_c - 2 \nabla_a \phi \nabla_b + 
g_{ab}(\partial \phi)^2 - 
2 \nabla_a \phi \nabla_b \phi
\right]_{(x)} R(y)
{1\over{\Box}}\nonumber \\
&+& {1\over{8\pi}} \int d^2x \sqrt{g}\int d^2y \sqrt{g}~\left\{
2 \partial_a {P(x)\over{\Box}} \partial_b {R(y)\over{\Box}}
- g_{ab} \partial^c{P(x)\over{\Box}}
\partial_c{R(y)\over{\Box}}\right\}\ .
\label{pepe}
\end{eqnarray}
Due to the presence of a term proportional to $\mbox{log} P$,
$\langle T_{ab}^{\rm I} \rangle $ given by Eq. (\ref{pepe}) has a singularity
when $P \rightarrow 0$. In particular, this seems to imply that $\langle 
T_{ab}^{\rm I} \rangle $
is singular even in Minkowski space, where $P\equiv 0$.
To elude this problem,  the authors
of \cite{balbi}  proposed a different energy-momentum tensor
defined {\it ad hoc}.
However, we would like to stress that this singularity is an artifact of the 
no-backscattering
approximation, since the effective action was obtained by assuming that $P$ 
has a non-zero constant value. In situations where $P\cong 0$, the 
no-backscattering approximation breaks down, and it is
more appropriate to use the effective action derived by expanding in powers 
of $P$, where
no pathology appears at $P=0$.
The origin of the $P=0$ singularity in Eq.~(\ref{pepe}) is quite clear:
it  is associated with the infrared divergence produced by the 
two-dimensional field $\psi$ which becomes massless when $P = 0$. In the 
approximation
in powers of $P$ (just as in the case of a free scalar field), this is 
regularized by the infrared cut-off $\mu $.

In both approximations the energy-momentum tensor 
is not conserved. Indeed, using the expansion in powers of $P$,
it is easy to check that 
the divergence of the energy momentum tensor is non-zero:
\begin{eqnarray}\nabla^b \langle T_{ab}\rangle =
&&{1\over{8\pi}}\left[\nabla^b \phi \nabla_a\nabla_b- 
\nabla^b \nabla_a \phi \nabla_b  - 2 \nabla_a \phi \Box 
- \nabla^b\nabla_a\phi\nabla_b\phi - \nabla_a\phi\Box\phi
\right]_{(x)}\nonumber \\
&& \times \int d^2y 
\sqrt{g}\left\{ R(y) 
{1\over{\Box}}- 2 {1\over{\Box}} 
\ln {-\Box\over{\mu^2}} P(y) + 2 \int d^2z \sqrt{g} {1\over{\Box}}
R(z){1\over{\Box}} P(y)\right\}.\label{diveravra}\end{eqnarray}
On the other hand, in the no-backscattering approximation
we have
\begin{eqnarray}\nabla^b \langle T_{ab}\rangle  &=& -{1\over{8\pi}} 
\nabla_a P + {1\over{8\pi}}\left[\nabla^b \phi \nabla_a\nabla_b- 
\nabla^b \nabla_a \phi \nabla_b\right. \nonumber \\
&&\left. - 2 \nabla_a \phi \Box - \nabla^b\nabla_a\phi\nabla_b\phi
- \nabla_a\phi\Box\phi
\right]~\mbox{log}{P\over{\mu^2}}.\label{divermukha}
\end{eqnarray}
As in the classical case Eq.(\ref{divt2d}),
the stress tensor is not conserved when the dilaton
field is not  constant.

It will be shown below that if the energy-momentum
tensor is computed by neglecting the invariant part of the effective 
action, so that $\langle T_{ab}\rangle = \langle T_{ab}^{\rm A} \rangle $,
one obtains wrong results for quantum effects in 
black hole and cosmological metrics. The same happens if 
$\langle T_{ab}\rangle $
is determined from the trace anomaly by imposing a conservation law.

\section{Hawking radiation}

\def\ov{\over}
\def\ha{{1\over 2}}
The Hawking radiation for a Schwarzschild black hole formed by gravitational 
collapse starting from the vacuum has been computed in \cite{fv} and recently 
discussed in \cite{balbi}.
The calculation can be easily extended to more general (asymptotically 
flat) backgrounds.
Let 
us consider the case of a general black hole,
formed by gravitational collapse of a shock wave at $v=v_0$. For 
$v<v_0$, 
the geometry is given by the Minkowski metric, i.e.
\begin{equation}
ds^2_{\rm in}=-du_{\rm in}dv_{\rm in}\ ,\ \ \ \ u_{\rm in}=t-r\ ,\ \ \ 
v_{\rm in}=t+r\ .
\end{equation}
For $v>v_0$, the geometry is 
\begin{equation}
ds^2=-\lambda(r) dudv  ,
\end{equation}
$$
u=t-r^*\ ,\ \ \  v=t+r^*\ ,\ \ \ \ {dr\ov dr^*}=\lambda(r) \ ,
$$
where $\lambda (r) $ vanishes at the event horizon $r=r_+$. For 
example, for a 
Reissner-Nordstr\" om black hole $\lambda(r)=1-{2M\ov r} +{q^2\ov r^2}$. 
The relation between ``in" and ``out" coordinates follows by matching the 
geometries at the infalling line $v=v_0$:
\begin{equation}
v=v_{\rm in}\ ,\ \ \ \ {du_{\rm in}\ov du}= \lambda\big(\ha(v_0-u_{\rm in})
\big)\ .
\label{inout}
\end{equation}
Let us first assume that $
P(r)={1\ov r}\lambda'(r)$
is different from zero everywhere outside the event horizon 
$r>r_+$ (this is the case for non extremal black holes).
We can therefore use the no-backscattering 
approximation. Adding 
Eqs. (\ref{niWefacc}) and (\ref{mukha}) the complete effective action reads
\begin{equation}
S_{\rm eff}= -{1\over 96\pi}\int d^2x \sqrt{g} ~R{1\over\Box}R + 
\mbox{local terms},
\label{caca}
\end{equation}  
i.e., up to local terms this effective action coincides with the 
ones for uncoupled scalar fields. 
In the calculation of Hawking radiation, only non-local terms  in
the effective action are relevant.

The four-dimensional energy-momentum tensor is given by
\begin{eqnarray}
\langle  T^{(4)}_{ab}\rangle &=&{1\over 2\pi}e^{2\phi}{1\over\sqrt g}
{\delta S_{\rm eff}\over
\delta g^{ab}} \ ,     \nonumber\\
\langle  T^{(4)}_{ij}\rangle &=&{1\over 8\pi}e^{2\phi}{1\over\sqrt g}
{\delta S_{\rm eff}\over
\delta \phi} g_{ij}\ ,
\label{4dt}
\end{eqnarray}
where the $i$ and $j$ indices denote the angular coordinates.
The information about Hawking radiation is contained in 
the components $\langle  T^{(4)}_{ab}\rangle $,
which  are 
in turn determined by the two dimensional energy momentum tensor
$\langle T_{ab}\rangle $. From Eq.~(\ref{caca}), and dropping the variation 
of the local terms, we have 
\begin{eqnarray}
\langle T_{ab}\rangle &=& -{1\over{24\pi}}  
\int d^2y \sqrt{g} \left[ \nabla_a \nabla_b - 
g_{ab} \Box \right]_{(x)} R(y) {1\over{\Box}} \nonumber \\
&+& {1\over{96\pi}} \int d^2x \sqrt{g}\int d^2y \sqrt{g}~\left\{
- g_{ab}\partial^c{R(x)\over{\Box}}\partial_c{R(y)\over{\Box}}
+ 2 \partial_a {R(x)\over{\Box}} \partial_b {R(y)\over{\Box}}\right\},
\label{f-v}\end{eqnarray}
and only the last term contributes to the Hawking radiation
\cite{fv}.
The formal expression ${1\over{\Box}} R $ in the equation above denotes the
retarded propagator $G_{ret}$ acting on the Ricci scalar.

In the conformal gauge $ds^2=-e^{2\rho}dx^+dx^-$ we have $-2\Box 
\rho = R$, therefore $\rho$ is formally given by $-2 \rho = {1\over{\Box}} 
R$. The retarded propagator
gives $-2 \rho_{\rm in}=G_{ret} R$ where $\rho_{\rm in}$ is one half the 
logarithm of the
scale factor in the ``in'' coordinates. The relation between the ``in''
and ``out'' scale factors is
\begin{equation}
e^{2\rho_{\rm in}}=e^{2\rho_{\rm out}}{du\over du_{\rm in}}{dv\over 
dv_{\rm in}}=e^{2\rho_{\rm out}}{du\over du_{\rm in}}.
\label{rhoinout}
\end{equation}

The energy flux through $I^+$ is given by
\begin{equation}
\langle T_{uu}\rangle_{I^+}=-{1\over{12\pi}}\left[{\partial^2\rho_{\rm in}
\over\partial u^2}-\left(
{\partial\rho_{\rm in}\over\partial u}\right)^2\right]_{I^+}.
\label{flujo}
\end{equation}
Using Eqs. (\ref{rhoinout}) and (\ref{inout}) we obtain
$$
2[\rho_{\rm in}]_{I^+}=\log {du\over du_{\rm in}}+ \mbox{const} =
-\log [\lambda ({1\over 2}(v_0-u_{\rm in})] + \mbox{const}
.$$
Combining the above equations we obtain
\begin{equation}
\langle T_{uu }\rangle_{I^+} ={1\ov 192\pi}
\lambda '^2\,\,\, ,
\label{zxxz}
\end{equation}
where $\lambda '=\lambda '(r_+)$.
This flux corresponds to a temperature 
($\langle T_{uu}\rangle ={\pi\ov 12} T_H^2$)

\begin{equation}
T_H={1\ov 4\pi}\lambda'(r_+)
.\label{temp}\end{equation}

Note that the above derivation applies for any 
asymptotically flat
black hole with metric 
$ds^2=-\lambda(r) dt^2+\lambda^{-1}(r) dr^2$. Indeed, the Hawking 
temperature for  a generic black hole
of this form 
(as obtained by going to Euclidean space and compactifying the time direction)
is given by  $T_H={1\ov 4\pi}\lambda'(r_{\rm hor})$,  in agreement with 
the flux (\ref{zxxz})
obtained above.

The lesson from this calculation is that, as long as $P$ is different
from zero, we can apply the no-backscattering approximation in order 
to compute Hawking radiation. The main contribution comes from the 
Polyakov term in the effective action and the result for the 
Hawking temperature
agrees with the well known four dimensional expression. The next 
to leading order contribution can be computed as described in Section V of
Ref. \cite{muk}.

Let us now consider a background geometry such that $P$ vanishes at the 
horizon, 
as is the case  for the Reissner-Nordstr\"om black holes in the extremal 
limit. In this 
situation the no-backscattering approximation still gives the correct result
for the Hawking radiation. Moreover, although $P$ vanishes at the horizon,
it is easy to check from  Eq. (\ref{pepe}) that there is no divergence in 
the energy momentum tensor. 
Alternatively, one can compute the Hawking
radiation for extremal black holes using  
the expansion in 
powers of $P$. Near the horizon the leading contribution in 
Eq. (\ref{niWefacc})  
is given by the 
non local
Polyakov term. Therefore the Hawking temperature is, to leading
order in $P$, again given by Eq. (\ref{temp}). 

It is important to stress
that the expansion in powers of $P$ is not useful to compute the
Hawking radiation for Schwarzschild black holes. Indeed, for this
geometry $P$ and $R$ are of the same order of magnitude, and one 
should add the contribution of an infinite number of 
non-local terms in order to 
obtain the correct radiation.

As a final remark, we stress that
if the Weyl invariant part $S_{\rm eff}^{\rm I}$ is 
neglected, the relevant terms for the Hawking radiation are
(see Eq. (\ref{ancon}))
\begin{equation}
{1\over{48\pi}} \int d^2x \sqrt{g}\int d^2y \sqrt{g}~\left\{
\partial_a {R(x)\over{\Box}} \partial_b {R(y)\over{\Box}}- 12
\partial_a {P(x)\over{\Box}} \partial_b {R(y)\over{\Box}}\right\}.
\end{equation}
Since $P=R/2$ for the Schwarzschild collapsing geometry, the term proportional 
to $P$ produces 
an infalling flux that exceeds by a factor $6$ the outgoing
one. Thus, if $S_{\rm eff}^{\rm I}$ is not taken into account, one would 
incorrectly obtain a negative 
energy-density flux of Hawking radiation. This problem appeared in 
\cite{muk,kummer}.

\section{Quantum correction to the Newtonian potential}

The different approximations can be tested by computing 
another observable: the quantum corrections to the Newtonian potential 
\cite{dalvit}.
The four-dimensional semiclassical Einstein equations read
\begin{equation}{1\over{8\pi}}(R_{\mu\nu}-{1\over{2}}g_{\mu\nu} R) 
= {}^{\rm class}T_{\mu\nu}^{(4)} 
+ \langle T_{\mu\nu}^{(4)}\rangle \ ,
\label{ee}
\end{equation} 
where ${}^{\rm class}T_{\mu\nu}^{(4)}$ is the four-dimensional classical 
contribution of a point particle of mass $M$, ${}^{\rm class}T_{\mu\nu}^{(4)} 
= - \delta^0_\mu \delta^0_\nu M \delta^3 (\vec x)$ and $\langle T_{\mu
\nu}^{(4)} \rangle$ is the energy-momentum tensor for a quantum massless 
scalar field. 

To solve these equations we consider perturbations around the flat spacetime 
$g_{\mu\nu} = \eta_{\mu\nu} + h_{\mu\nu}$.
For our purposes it is enough to solve the equation for the trace of 
$h_{\mu\nu}$ to find the quantum corrections. In a perturbative expansion, 
$h = h^{(0)} + h^{(1)}$, with 
$ h^{(0)}={4 M\over r}$ coming from the classical solution. 
The equation for $h^{(1)}$
is 
\begin{equation}
{1\over{2\pi}}\nabla^2 h^{(1)} = 
g^{\mu\nu}\langle T_{\mu\nu}^{(4)}\rangle \ .
\end{equation}
At large 
distances the trace of $\langle T_{\mu\nu}^{(4)}\rangle$ is given by 
\cite{dalvit}
\begin{equation}
\langle T^{(4)}\rangle =-{M\over 8\pi^2 r^5} \equiv {C\over r^5}\ .
\label{ffz}
\end{equation}
The perturbative solution to the semiclassical Einstein equations is
therefore
\begin{equation}-{h\over 4} = - {M\over r}+ {M\over{12\pi}}
 {1\over{r^3}} + ....\end{equation}
from which it is possible to read the quantum corrections to the Newtonian
potential. 
For a minimally coupled massless 4D scalar
 the stress tensor trace is state dependent. Equation (38)
corresponds to computing the trace of the stress tensor in the 
Boulware state. 
The expression (38) is in agreement with other calculations 
of quantum corrections to the Newton potential \cite{varios}.
This term  seems to be however missing in the treatment of 
ref.~\cite{anderson}.
In this work a comparison of numerical and analytic results was made only 
near the horizon.
A complete treatment valid at large distances as well must give a trace 
of the four dimensional 
energy momentum tensor proportional to ${M\over r^5}$ as $r\rightarrow\infty$
in order to reproduce the correct answer for the quantum corrected
potential \cite{varios}.

{}From the analysis above we see that in order to compute the leading quantum
corrections it is necessary to evaluate the (four-dimensional) trace of the
energy momentum tensor in the Schwarzschild background. 
It is interesting to compute it now in the reduced model Eq.~(\ref{2dact}).
On general grounds we expect 
$g^{\mu\nu}\langle T_{\mu\nu}^{(4)}\rangle ={C\over r^5}$ where 
$C = C(\mu r)$. The 
sign of $C$ is very 
important. Indeed, a negative value of $C$ implies that the Newton constant
grows with $r$, as suggested by the fact that there is no screening of 
the gravitational interaction by quantum matter fields.

The no-backscattering approximation is 
not adequate to describe the vacuum polarization in the asymptotically
flat region. Indeed, from Eq. (\ref{pepe}) we see that for the Schwarzschild
metric the energy momentum tensor contains terms proportional to
${1\over r^2}\ln ({M\over \mu^2 r^3})$ as $r\rightarrow \infty$. These
do not vanish (in fact diverge \cite{balbi}) as $M\rightarrow 0$.
Therefore, the four-dimensional 
trace $\langle T^{(4)} \rangle =
g^{\mu\nu}\langle T_{\mu\nu}^{(4)}\rangle =
g^{ab}\langle T_{ab}^{(4)}\rangle + g^{ij}\langle T_{ij}^{(4)}\rangle$
must be computed using the expansion in powers of $P$ for the effective 
action. In this approximation we must evaluate Eqs.~(\ref{ancon}) and 
(\ref{nanoool}) in the collapsing metric 

\begin{equation}ds^2 = \left(1 - {2M\over{r}}\right)\left(-dt^2 + 
dr^{\star 2}\right) + r^2 d\Omega^2,\label{schmet}\end{equation} 
where $d\Omega^2$ is the line element of the unit two-sphere, and $r^{\star}$
is given by
\begin{equation}r^{\star} = r + 2 M \ln{\vert {r\over{2M}} - 1\vert}
.\end{equation}
In this metric $R={4M\over{r^3}}$ and $P = {R\over{2}}$. The non-local 
functions ${R\over{\Box}}$ and  ${P\over{\Box}}\ln {{-\Box\over{\mu^2}}}$
 are computed by means of their Fourier transforms \cite{lith}, and they 
are given by    

$${R\over
{\Box}} = {2M\over{r}} \\\\~~~~  \mbox{and} ~~~~ \\\\  {P\over{\Box}}\ln 
{{-\Box\over{\mu^2}}} =  -{2M\over{r}}\ln {{\tilde \mu} r}.$$ 

The four dimensional components of the energy-momentum tensor are 
given by Eqs. (\ref{4dt}).  
Evaluating Eqs. (\ref{ancon}), (\ref{nanoool}), and  
taking the functional variation with respect to the dilaton field, we obtain 
the four dimensional trace, up to linear 
order in $M$:

\begin{equation}
\langle T^{(4)}\rangle = - {1\over{8 \pi^2}}{M\over{r^5}} 
\ln {{\tilde \mu} r}\ .
\end{equation}
As expected, quantum corrections to the 
Newtonian potential depend on $\mu$. This correction agrees qualitatively 
with the four-dimensional result (\ref{ffz}), i.e.,
it has the correct sign. However, if the Weyl invariant part of
the effective action were neglected, one would obtain
\begin{equation}
\langle T^{(4)}\rangle = {1\over{48 \pi^2}}{M\over{r^5}}\ ,
\end{equation}
which would lead to quantum corrections to the Newtonian potential with 
the wrong sign.

\section{Cosmological particle creation}

As another example, in this Section we consider  
particle creation in cosmological backgrounds. 
Let us consider the metric

\begin{equation}ds^2 = a^2(t)[-dt^2 + dr^2] + a^2(t) r^2 d\Omega_2,
\end{equation}
where $a(t) = 1 + \delta (t) $ with $\delta << 1$ and $\delta 
\rightarrow 0$ in the far past and future. We denote by $t$ the conformal 
time. 

The total number of created particles is given by the imaginary part of 
the in-out effective action. This effective action 
can be obtained from the Euclidean effective action replacing the Euclidean 
propagators by the Feynman ones. As $P \approx \ddot \delta$, the 
approximation 
in powers of $P$ is adequate in order to evaluate particle creation rate. Up 
to lowest order 
in $\delta$, the Euclidean effective action is given 
by Eqs. (\ref{niWefacc}) and (\ref{avra}), where the propagators are 
the flat spacetime ones. 

In the conformal vacuum the terms present in the anomalous part of the 
effective action ($S_{\rm eff}^{\rm A}$) 
are real and local for cosmological metrics. The 
invariant part  $S_{\rm eff}^{\rm I}$ is non-local and contains an 
imaginary term that gives the particle creation.

Performing a Fourier transform of Eq. (\ref{avra}), and replacing 
$p^2 \rightarrow p^2 - i \epsilon$ we obtain 
 
\begin{equation}
S_{\rm eff}^{\rm in-out} = {1\over{16\pi^2}}\int d^2p \vert 
{\tilde P}(p)\vert^2
{1\over{p^2 - i \epsilon}} \ln{{p^2 - i \epsilon}\over{\mu^2}} + 
\mbox{local terms}.
\end{equation} 
Using the fact that 

\begin{equation}
\ln{{p^2 - i \epsilon}\over{\mu^2}} = 
\ln \vert {{p^2}\over{\mu^2}}\vert - i \pi \theta (-p^2),
\end{equation}
the total number of created particles is given by

\begin{equation}
n_{\rm T} = {\rm Im} S_{\rm eff}^{\rm in-out} = - {1\over{16 \pi}} \int d^2p 
\vert {\tilde P}(p)\vert^2 {\theta (-p^2)\over{p^2}}.
\label{nt}
\end{equation}

Since $P = P(t)$, $n_{\rm T}$ takes the form

\begin{equation}
n_T = {\rm Im} S_{\rm eff}^{\rm in-out} = {1\over{16 V \pi}} \int dp_0 
\vert {\tilde P}(p_0)\vert^2 {1\over{p_0^2}},
\label{zzz}
\end{equation}
where $V$ is the spatial volume.

Because the metric is asymptotically flat for $t\rightarrow \pm \infty$, 
the Fourier transform ${\tilde P}(p_0)$ vanishes
as  $p_0 \rightarrow 0$. As a result,  the total number of created particles 
$n_T$ given in Eq.~(\ref{zzz}) is a finite quantity.

Equation~(\ref{nt}) represents  the precise two-dimensional 
analogue of the general expression for the total number of created particles 
in four dimensions (in the case of $\xi = 0$, 
$m = 0$, and $C_{abcd} = 0$) given in Ref. \cite{frie}.

It is important to note that the effective action coming from the 
no-backscattering approximation (\ref{caca}) is not adequate to evaluate the 
particle creation rate because the Polyakov term becomes real and local 
in the conformal vacuum. This would imply vanishing particle creation, in 
contradiction with the four dimensional result.

\section{Final remarks}

To summarize, we have shown that the Weyl invariant part of the effective 
action
contains relevant information about quantum effects in black hole geometries. 
Neither the effective action nor the mean value of the energy
momentum tensor can be completely determined
by the trace anomaly when the matter fields are coupled to the dilaton.
Neglecting this term, or imposing the conservation law for the two-dimensional
energy-momentum tensor, leads to wrong results for black hole 
radiation,   
quantum corrections to the Newtonian
potential and   cosmological particle creation.

We have discussed two different approximations in order to compute the 
invariant part of the effective action: the no-backscattering 
approximation introduced in Ref. \cite{muk}, and an expansion in powers of 
$P$. The no-backscattering 
approximation assumes a constant, non-zero value of $P$, and can be 
improved by 
performing an expansion in powers of derivatives of $P$ around this 
non-zero value. This was made in sect.~V of  \cite{muk},
where a ``backscattering" part of the effective action was added to the
no-backscattering part to get the total s-channel effective action.
 One expects the no-backscattering 
approximation to be valid for $P^2 \gg \nabla\nabla P$,  and therefore it 
is not applicable for the evaluation of the 
mean value of the energy momentum tensor for nearly flat metrics. However, it 
is adequate in order 
to determine the Hawking flux of black holes. 

On the 
other hand, the expansion in powers of $P$ is adequate in situations where 
$P^2 \ll \nabla\nabla P$, such as nearly degenerate Reissner-Nordstr\"om black 
holes, or to evaluate $\langle T_{ab}(r) \rangle$ outside a star whose 
radius $R$ is such that $R>2M$. Therefore it is useful to 
compute quantum corrections to the Newtonian potential. It is also useful
to compute cosmological particle creation for weak gravitational 
fields. In this 
approximation, the results depend on an infrared cut-off that appears 
because the model contains massless fields in two dimensions.

\acknowledgments
We would like to thank R. Bousso and A. Fabbri for useful comments.
The work of F.L. and F.M. was supported by Universidad de Buenos Aires, 
Conicet and Fundaci\'on Antorchas. F.D.M. would like to thank the British
Council and the hospitality
of the theoretical group at Imperial College, where part of this work 
has been done.  The work of J.R is supported by the European
Commission TMR programme grant  ERBFMBI-CT96-0982.

\end{document}